\documentclass[12pt,preprint]{aastex}

\begin{document}

\title{On the Formation of Gas Giant Planets on Wide Orbits}

\author{Alan P.~Boss}
\affil{Department of Terrestrial Magnetism, Carnegie Institution of
Washington, 5241 Broad Branch Road, NW, Washington, DC 20015-1305}
\authoremail{boss@dtm.ciw.edu}

\begin{abstract}

A new suite of three dimensional radiative, gravitational hydrodynamical
models is used to show that gas giant planets are unlikely to form 
by the disk instability mechanism at distances of $\sim$ 100 AU to
$\sim$ 200 AU from young stars. A similar result seems to hold for 
the core accretion mechanism. These results appear to be consistent 
with the paucity of detections of gas giant planets on wide orbits
by infrared imaging surveys, and also imply that if the object orbiting
GQ Lupus is a gas giant planet, it most likely did not form at
a separation of $\sim 100$ AU. Instead, a wide planet around GQ Lup
must have undergone a close encounter with a third body that tossed 
the planet outward to its present distance from its protostar. If it 
exists, the third body may be detectable by NASA's 
{\it Space Interferometry Mission}.

\end{abstract}

\keywords{stars: planetary systems -- stars: low-mass, brown dwarfs}

\section{Introduction}

 Because radial velocity planet-finding surveys are most sensitive
to planets with relatively short-period orbits, the search for
gas giant planets on wide orbits (i.e., separations much greater than
$\sim 10$ AU) has been undertaken primarily by direct imaging surveys
at infrared wavelengths. These surveys have largely turned up
empty-handed: McCarthy \& Zuckerman (2004) found no evidence for
any planets with masses of 5 to 10 $M_J$ (Jupiter mass) at distances of
75 to 300 AU from $\sim$ 100 stars. Similarly, Masciadri et al. (2005)
found nothing in their search of 28 stars for similar-mass planets 
with separations of at least $\sim 36$ to $\sim 65$ AU, as did
Lowrance et al. (2005) in their search of 45 nearby stars.
 
 However, recently two ground-based surveys have detected
very low mass companions with wide separations. Using the NACO
adaptive optics systems on the Very Large Telescope, 
Chauvin et al. (2004) found a $\sim 5 M_J$ companion
to the $\sim 25 M_J$ brown dwarf 2M1207, with a separation of
$\sim 60$ AU. Neuhauser et al. (2005) used
the same NACO system to detect an object orbiting $\sim 100$ AU 
from the $\sim$ 1 Myr-old classical T Tauri star GQ Lup. The mass of this
object appears to lie between $\sim 1 M_J$ and $\sim 42 M_J$,
with a good chance that its mass is low enough ($< 13 M_J$)
for it to be classified as a planet rather than as a brown dwarf.

 Core accretion (Mizuno 1980), the generally accepted mechanism 
of gas giant planet formation, encounters difficulties in forming
gas giants at distances much greater than $\sim 5$ AU (Pollack et al. 
1996; Inaba, Wetherill, \& Ikoma 2003) because of the decreasing
surface density of solids available to make the solid cores and
the consequent increase in the formation time scale. Forming gas 
giant planets at distances of $\sim 100$ AU by core accretion
appears to be extremely unlikely.

 Observations of the circumstellar disk orbiting the Herbig Ae star
AB Aurigae show clear evidence for spiral structures at both near-infrared
(Fukagawa et al. 2004) and millimeter-wavelengths (Corder, Eisner, \&
Sargent 2005). Four trailing spiral arms can be resolved, at distances 
of 200 AU to 450 AU from the central $\sim$ 4 Myr-old star with a mass of
2.8 $M_\odot$. These observations are perhaps the first direct evidence
for large scale gravitational instability in a circumstellar disk,
and suggest that the competing mechanism for gas giant planet formation, 
disk gravitational instability (Cameron 1978; Boss 1997, 2003), might
be able to operate at distances of $\sim 100$ AU or more. This paper
examines the latter possibility, by extending the same disk instability
models that suggest gas giant planet formation is possible at 
distances of $\sim 10$ AU, to examine the case of much larger 
disks, $\sim 100$ AU in size.

\section{Numerical Methods}

 The calculations were performed with a finite volume code
that solves the three dimensional equations of hydrodynamics and
radiative transfer, as well as the Poisson equation for the gravitational 
potential. The code is second-order-accurate in both space and time 
(Boss \& Myhill 1992) and has been used extensively in previous 
disk instability studies (e.g., Boss 2003). 

 The equations are solved on spherical coordinate grids with 
$N_r = 101$, $N_\theta = 23$ in $\pi/2 \ge \theta \ge 0$, 
and $N_\phi = 256$ or 512. The radial grid extends from 
100 AU to 200 AU with a uniform spacing of $\Delta r = 1$ AU.
The $\theta$ grid is compressed toward the midplane in order to ensure 
adequate vertical resolution ($\Delta \theta = 0.3^o$ at the midplane). 
The $\phi$ grid is uniformly spaced to prevent any azimuthal bias. 
The central protostar wobbles in response to the growth of 
disk nonaxisymmetry, preserving the location of the center 
of mass of the star and disk system. The number of terms in the 
spherical harmonic expansion for the gravitational potential of the disk
is $N_{Ylm} = 32$ or 48. The Jeans length criterion is monitored
throughout the calculations to ensure proper spatial resolution: the 
numerical grid spacings in all three coordinate directions always 
remain less than 1/4 of the local Jeans length. 
 
 The boundary conditions are chosen at both 100 AU and 200 AU to absorb radial 
velocity perturbations. Mass and linear or angular momentum entering 
the innermost shell of cells at 100 AU is added to the central protostar
and thereby removed from the hydrodynamical grid. Similarly,
mass and momentum that reaches the outermost shell of cells at 200 AU
is effectively removed from the calculation: the mass piles up in this
shell and is assigned zero radial velocity. The inner and outer boundary
conditions are designed to absorb incident mass and momentum, rather
than to reflect mass and momentum back into the main grid. The angular momentum
added to the central protostar is used only to monitor the conservation
of total angular momentum during the calculation. 

 As in Boss (2003), two of the models treat radiative transfer in the
diffusion approximation, with no radiative losses or gains occuring in 
regions where the vertical optical depth $\tau$ drops below 10.
In very low density regions, the disk temperature is assumed to
be the same as that of the disk envelope, 30K. The energy equation 
is solved explicitly in conservation law form, as are the four 
other hydrodynamic equations. 

\section{Initial Conditions}

 The models calculate the evolution of a $1 M_\odot$ 
central protostar surrounded by a protoplanetary disk with 
a mass of 0.16 $M_\odot$ between 100 AU and 200 AU. 
The models envision planet formation as occurring during the
embedded phase of star formation, when the star is a Class I
object still accreting mass from the infalling cloud envelope
onto a relatively massive protoplanetary disk. Given that
the disk mass interior to 100 AU would add another $\sim 0.1
M_\odot$, the total amount of circumstellar matter is assumed 
to be perhaps unrealistically high, making the negative results
regarding forming gas giants by disk instability at such
large distances obtained here even stronger than would be
the case for an assumed lower mass disk system.

 The initial protoplanetary disk structure is based on
the following approximate vertical density distribution (Boss 1993) for
an adiabatic, self-gravitating disk of arbitrary thickness in
near-Keplerian rotation about a point mass $M_s$

$$ \rho(R,Z)^{\gamma-1} = \rho_o(R)^{\gamma-1} $$
$$ - \biggl( { \gamma - 1 \over \gamma } \biggr) \biggl[
\biggl( { 2 \pi G \sigma(R) \over K } \biggr) Z +
{ G M_s \over K } \biggl( { 1 \over R } - { 1 \over (R^2 + Z^2)^{1/2} }
\biggr ) \biggr], $$

\noindent where $R$ and $Z$ are cylindrical coordinates, $\rho_o(R)$ 
is a specified midplane density, and $\sigma(R)$ is a specified
surface density. The disk's surface is defined by locations
where the density distribution defined above falls to zero;
regions where the density falls below zero are outside the 
disk. The adiabatic pressure (used only for defining
the initial model -- the radiative transfer solution includes a full
thermodynamical treatment) is defined by $p = K \rho^\gamma$,
where $K$ is the adiabatic constant and $\gamma$ is the adiabatic
exponent. The adiabatic constant is $K = 5.1 \times 10^{17}$ (cgs units)
and $\gamma = 5/3$ for the initial model. $K$ was chosen to be
three times larger than in the previous disk instability models
in order to produce a thicker outer disk. The radial
variation of the midplane density is a power law that ensures
near-Keplerian rotation throughout the disk

$$\rho_o(R) = \rho_{o4} \biggl( {R_4 \over R} \biggr)^{3/2}, $$

\noindent where $\rho_{o4} = 4.0 \times 10^{-11}$ g cm$^{-3}$ and
$R_4 = 4$ AU. With this assumption for $\rho_o(R)$, the initial disk 
surface density profile from 100 AU to 200 AU is $\sigma_i \propto r^{-2}$, 
a steeper falloff than occurs for this choice of an initial disk model
at radii of $\sim$ 20 AU, where $\sigma_i \propto r^{-3/2}$ (Boss 2002).
A low density halo $\rho_h$ of gas and dust infalls onto the disk, 
with

$$ \rho_h(r) = \rho_{h4} cos^2(\theta)
\biggl ( {R_{100} \over r} \biggr)^{3/2}, $$

\noindent where $\rho_{h4} = 4.5 \times 10^{-15}$ g cm$^{-3}$, 
$R_{100} = 100$ AU, and $r$ is the spherical coordinate radius. 
The initial envelope mass is 0.05 $M_\odot$.

Three initially uniform disk temperatures are investigated, $T_o = 20$, 
25, and 30K. With the assumed initial density profile, the 
disks have initial $Q$ gravitational stability parameters as low
as $Q_{min} = 1.3$ for $T_o = 20$K and $Q_{min} = 1.6$ for $T_o = 30$K.
In low optical depth regions such as the disk envelope, 
the temperature is assumed to be 30 K. The Rosseland mean opacities
used in the radiative transfer solution have been updated to
include the dust grain opacities calculated by Pollack et al. (1994).

\section{Results}

 Table 1 lists the initial conditions for the models as well as
the final times ($t_f$) to which they were evolved. Note that
for the initial disk, the orbital period is 740 yrs at 100 AU
and 2400 yrs at 200 AU, so the $N_\phi = 256$ models were 
evolved for times that varied between $\sim$ 10 and $\sim$ 110
orbital periods at 100 AU, while the $N_\phi = 512$ models were 
calculated for between $\sim$ 5 and $\sim$ 27 orbital periods
at 100 AU. 

 Models A and AH began with a minimum $Q$ parameter of 1.30, and thus 
had a greater tendency toward instability than models C and CH, where
the initial minimum $Q$ value was 1.59, implying initial marginal
gravitational instability. Models B and BH were intermediate
with a value of 1.45. Models A and AH were evolved with the
usual radiative transfer procedures, but it was found to be necessary
for numerical stability to evolve the other four models (B, BH, C, CH) 
with the disk temperature fixed at its initial value. The latter
assumption errs on the side of encouraging clump formation, 
compared to the full thermodynamical treatment in models A and AH,
but we shall see that even this assumption does not result in
clump formation.

 Figure 1 shows the result of model AH after 4000 yrs of evolution.
A tightly wound set of spiral arms has formed, but no dense clumps
capable of contracting to become self-gravitating protoplanets
appear, nor do any appear during the earlier phases of the 
evolution. This is in spite of the fact that model AH is the model
that should be most likely to form clumps if clump formation
is possible, given its low initial minimum value of $Q = 1.30$
and its relatively high spatial resolution ($N_\phi = 512$).

 The main reason why clumps do not form in this model (as well
as the others) appears to be that rapid inward transport of mass 
associated with the gravitational torques between the growing spiral
arms is able to deplete the inner disk gas before it can become
dense enough to form gravitationally-bound clumps. This is evident 
in Figure 2, which shows how the disk in model AH has developed a severe 
inner depletion of gas that renders the disk locally gravitationally
stable by a large factor. The relatively steep decrease in
disk surface density at these distances ($\sigma_i \propto r^{-2}$)
means that the loss of the innermost disk gas drives the
disk toward stability.

 The tightness of the spiral arms in model AH is further
illustrated in Figure 3, which shows the equatorial temperature
contours at the same time as Figure 1. The spiral arms are
delineated by sharply defined, tightly wound temperature
maxima that occur where the disk gas is being compressed
locally. The fact that the spiral structures that form are
so tightly wound appears to also work against clump
formation, as this configuration prevents the superposition of
multiple spiral arms that would lead to locally higher gas densities,
a process that occurs frequently in disk instability models 
on scales of $\sim$ 10 AU (e.g., Boss 2003) and helps lead
to gravitationally-bound clump formation. The spiral
arms in Boss (2003) are loosely wound in comparison
to those seen in Figures 1 and 3. In the latter models, 
clumps formed within $\sim$ 300 yrs, or $\sim$ 3 orbital
periods at $\sim$ 20 AU.

 Given the failure of model AH to produce any dense clumps,
it should be obvious that in the remaining five models,
clump formation is even more inhibited by the same processes
that afflicted model AH. This is in spite of the fact that
in models B, BH, C, and CH, the disk was assumed to remain
isothermal during the evolution for numerical reasons, an
assumption that can be expected to err on the side of clump
formation. Figure 4 shows that in model CH,
with an initial minimum value of $Q = 1.59$ and $N_\phi = 512$,
the disk becomes remarkably axisymmetric after its inner
regions have been depleted by accretion onto the central
protostar.

\section{Conclusions}

 These models have shown that the disk instability mechanism has 
great difficulty forming gas giant planets {\it in situ} at distances
of $\sim$ 100 AU to $\sim$ 200 AU from solar-mass protostars.
The models presented here show no clump formation, even in 
an initially marginally gravitiationally unstable disk.
Given that the core accretion mechanism is even more hampered
in forming gas giant planets at $\sim$ 100 AU distances
(Pollack et al. 1996; Inaba, Wetherill \& Ikomoa 2003), it
seems clear that gas giant planets do not appear to be
able to form on such wide orbits, consistent with the failure
of most direct imaging surveys to detect massive gas giant planets
(McCarthy \& Zuckerman 2004; Masciadri et al. 2005; 
Lowrance et al. 2005).

 On the other hand, the detection of the wide, very low mass
objects around 2M1207 by Chauvin et al. (2004) and around 
GQ Lup by Neuhauser et al. (2005) demands an explanation. The
2M1207 system would seem to be best explained as a binary
system composed of a brown dwarf and a sub-brown dwarf,
with a mass ratio $q = 0.2$, similar to that of many
binary star systems. Binary and multiple star systems
are generally believed to have formed from the collapse
and fragmentation of dense molecular clouds, a process that
precedes the planet formation processes considered here.

 If the very low mass companion to GQ Lup turns out to
have a mass close to the upper limit of $\sim 42 M_J$,
then it also most likely formed as a brown dwarf companion
to GQ Lup through collapse and fragmentation into a binary 
system with $q \le 0.06$ (assuming a mass of $\sim 0.7 M_\odot$
for the T Tauri star GQ Lup). However, if the object has
a much lower mass, implying that it did not form by fragmentation,
but through the planet formation process in a protoplanetary
disk, it is hard to see how such an object could have formed
{\it in situ} at $\sim$ 100 AU by either core accretion or
disk instability. Instead, one would be forced to consider
a scenario where the object formed much closer to GQ Lup,
perhaps within $\sim$ 30 AU (e.g., Boss 2003), and then was
flung outward on a highly eccentric orbit to $\sim$ 100 AU
by a close encounter with another body in the GQ Lup system
(cf., Debes \& Sigurdsson 2006). Outward scattering of planets 
to wide orbits is a likely possibility in crowded systems of 
giant planets (Adams \& Laughlin 2003). 

 NASA's {\it Space Interferometry Mission (SIM)} should be able to
detect such a third body orbiting relatively close to GQ Lup,
if it exists, settling the question of the origin of the putative
planet's wide orbit. GQ Lup has a V magnitude of 14.4 but is
highly reddened, allowing {\it SIM} to perform narrow angle
astrometry on it in the R band. Because the third body would
be left on a tighter orbit around GQ Lup as a result of the close 
encounter with the planet now on a wide orbit, the third body would
be expected to be on a relatively short period orbit that could
be detected during {\it SIM's} nominal mission lifetime of 5 yrs.
At GQ Lup's distance of 140 pc, {\it SIM} would be able to
detect a third body with a mass as low as $\sim 1 M_J$ orbiting
with a semi-major axis of 10 AU or less. As the third body is
likely to be significantly more massive than the wide planet, 
{\it SIM} should be able to detect this hypothetical object.
 
 I thank Alycia Weinberger for motivating these calculations, 
Charles Beichman for advice about {\it SIM's} capabilities, 
Fred Adams for his excellent comments on the manuscript, and 
Sandy Keiser for her extraordinary computer systems expertise. This 
research was supported in part by NASA Planetary Geology and Geophysics
grant NNG05GH30G and by NASA Astrobiology Institute grant NCC2-1056.
The calculations were performed on the Carnegie Alpha Cluster, 
the purchase of which was partially supported by NSF Major Research
Instrumentation grant MRI-9976645.

\begin{figure}
\vspace{-2.0in}
\plotone{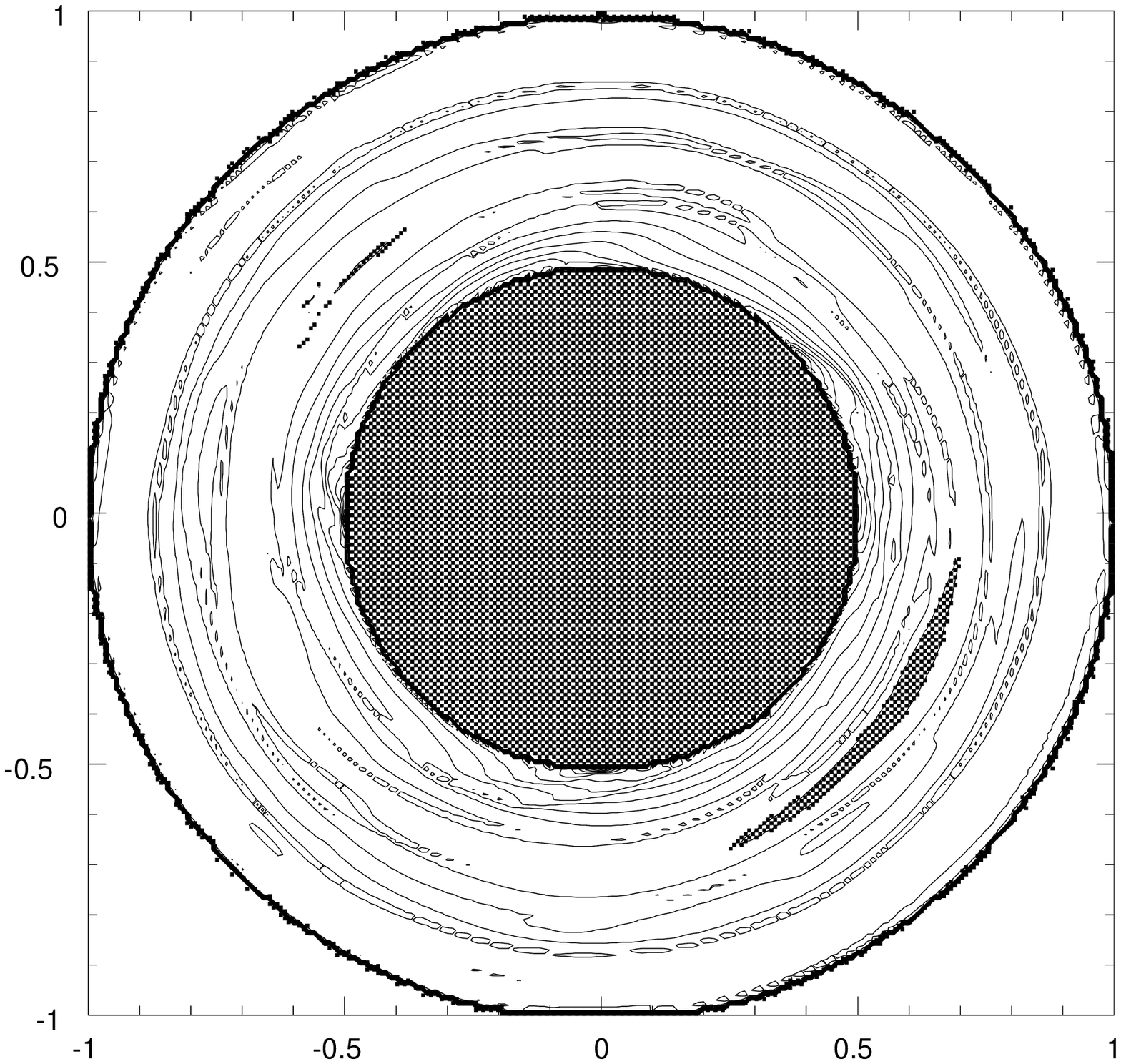}
\caption{Equatorial density contours for model AH after 4000 yrs of evolution. 
In this figure as well as in the subsequent figures, a $0.16 
M_\odot$ disk is in orbit around a central 1 $M_\odot$ protostar. The 
entire disk is shown, with an outer radius of 200 AU and an inner radius of 
100 AU, through which mass accretes onto the central protostar. Hashed regions 
denote spiral arms with densities higher than $3.2 \times 10^{-14}$ 
g cm$^{-3}$. Density contours represent factors of two change in density.}
\end{figure}

\begin{figure}
\vspace{-2.0in}
\plotone{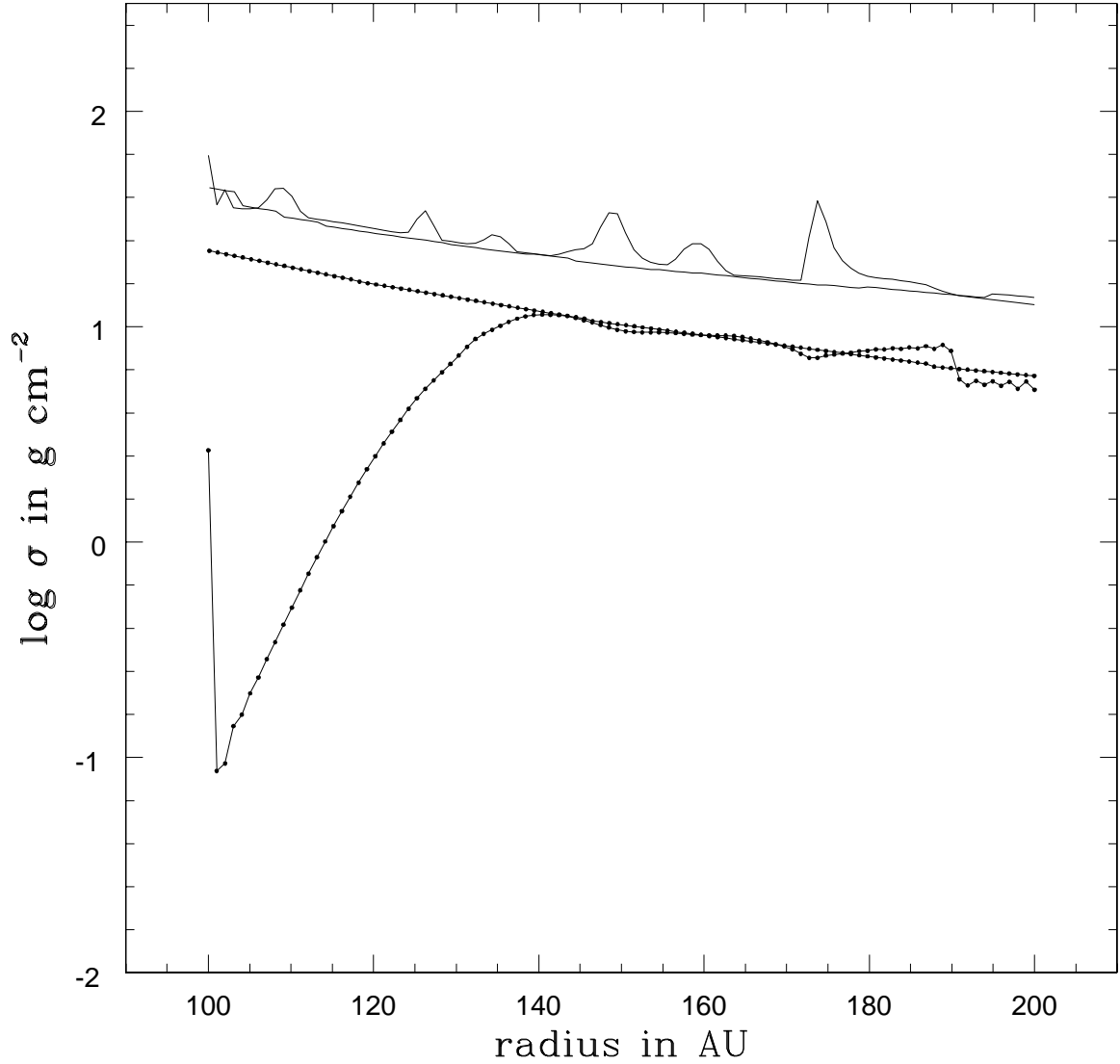}
\caption{Radial (azimuthally averaged) profiles for the disk gas surface 
density (dots) in the initial model AH and after 4000 yrs of evolution, 
compared to the surface densities needed for $Q = 1$ (solid line)
at both times. The smoother curves are the initial profiles for both
quantities. After 4000 yrs, the innermost disk gas has been largely 
accreted inside the inner 100 AU boundary.}
\end{figure}

\begin{figure}
\vspace{-2.0in}
\plotone{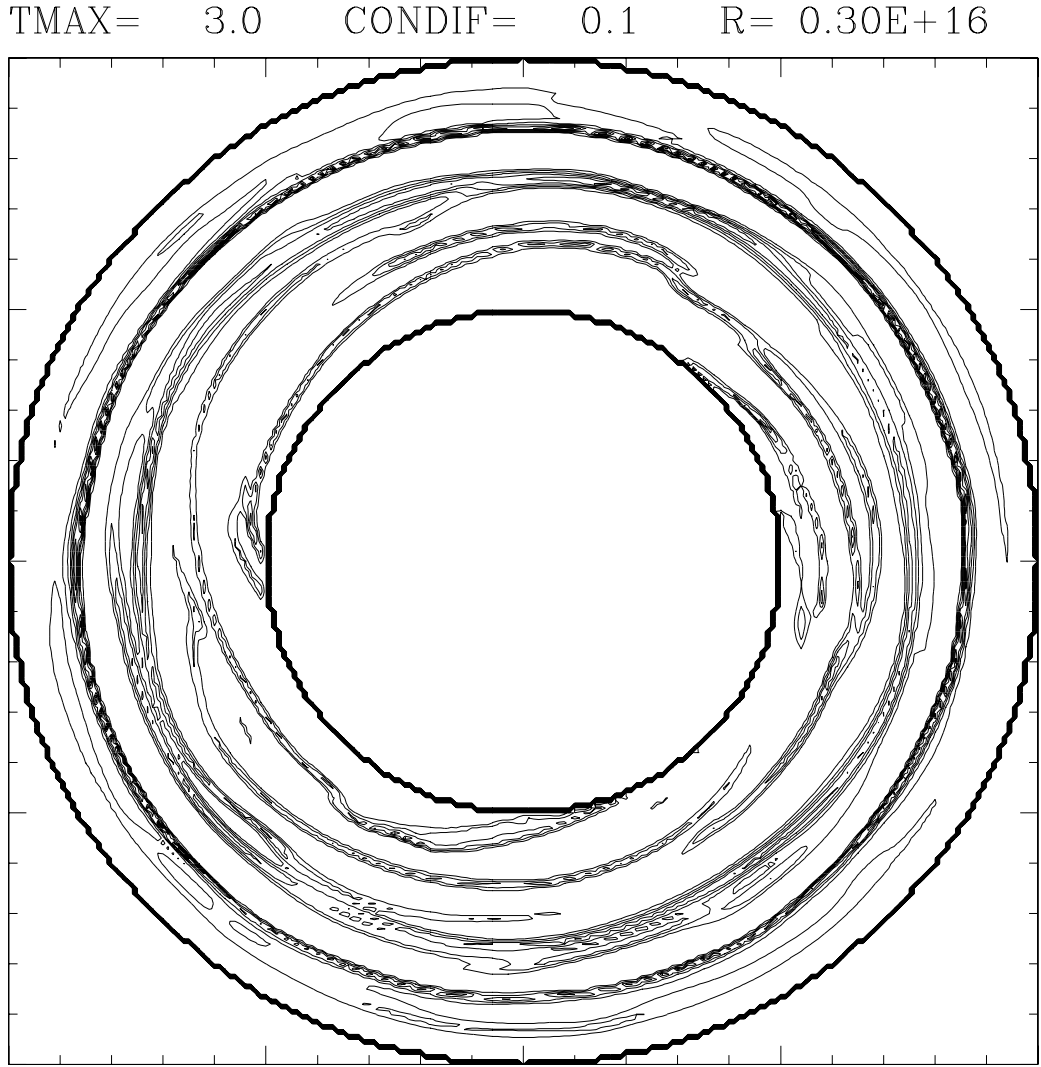}
\caption{Equatorial temperature contours for model AH after 4000 yrs of 
evolution, as in Figure 1. Temperature contours represent factors of 
1.3 change in temperature.}
\end{figure}

\begin{figure}
\vspace{-2.0in}
\plotone{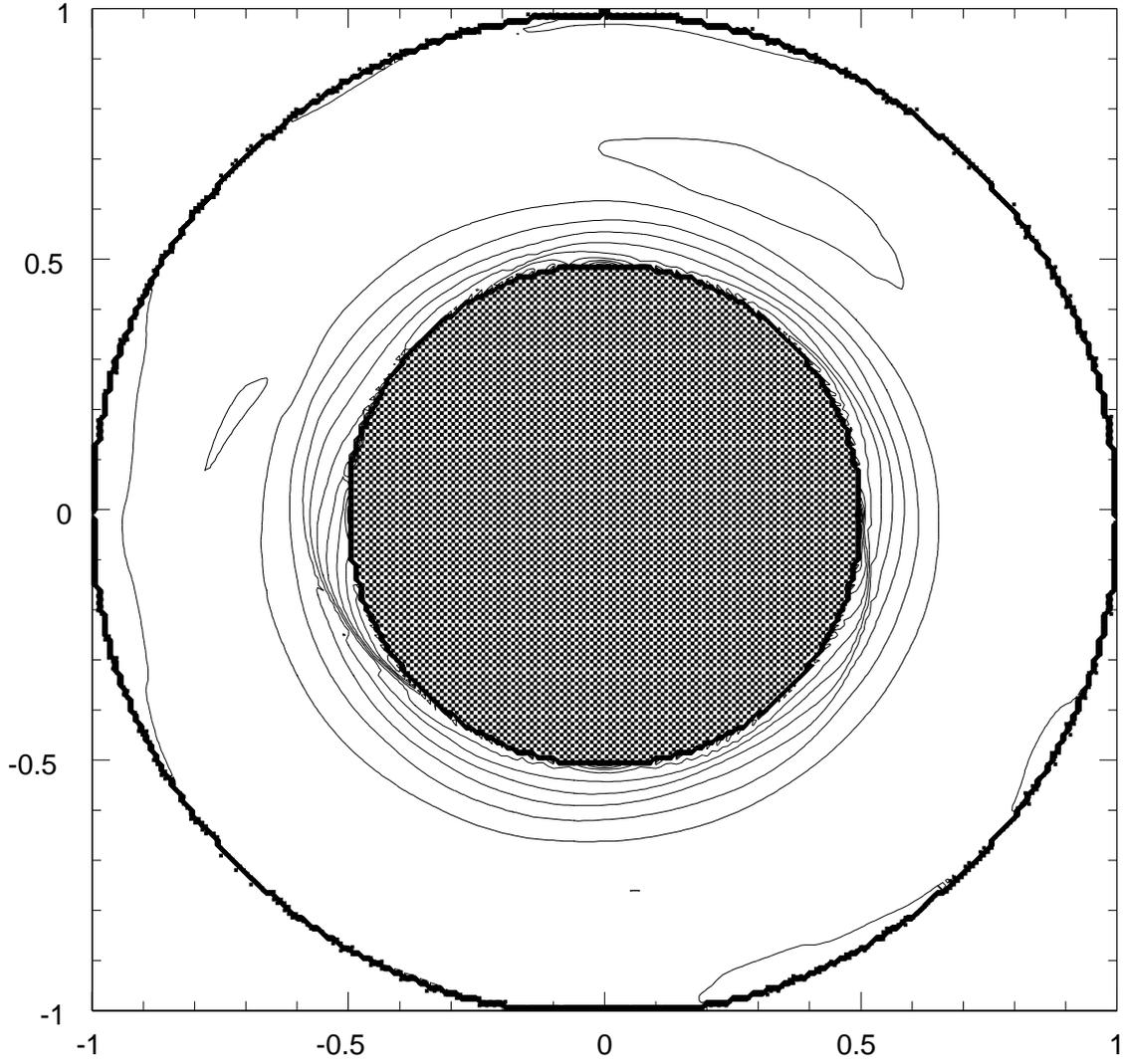}
\caption{Equatorial density contours for model CH after 20000 yrs 
of evolution, as in Figure 1.}
\end{figure}

\suppressfloats

\clearpage
\begin{deluxetable}{cccccccc}
\tablecaption{Initial conditions for the models.\label{tbl-1}}

\tablehead{\colhead{model} & 
\colhead{\quad \quad $T_o$ (K) \quad } & 
\colhead{\quad \quad $min(Q_i)$ \quad } & 
\colhead{\quad \quad $N_\phi$ \quad } & 
\colhead{\quad \quad $N_{Ylm}$ \quad } & 
\colhead{\quad \quad $t_f$ (yrs)} }

\startdata

A  & 20 & 1.30 & 256 & 32 & 7,800  \\

AH & 20 & 1.30 & 512 & 48 & 4,000  \\

B  & 25 & 1.45 & 256 & 32 & 35,000 \\

BH & 25 & 1.45 & 512 & 48 & 5,000  \\

C  & 30 & 1.59 & 256 & 32 & 84,000 \\

CH & 30 & 1.59 & 512 & 48 & 20,000 \\

\enddata
\end{deluxetable}
\clearpage

\suppressfloats

\end{document}